\documentstyle[twocolumn,aps,epsf]{revtex}
\newcommand{\ignore}[1]{}

\begin{document}
\twocolumn[
\hsize\textwidth\columnwidth\hsize\csname @twocolumnfalse\endcsname
\draft
\title{Time dependence of transmission in semiconductor superlattices} 
\author{C. N. Veenstra$^{a,b}$, W. van Dijk$^{a,b}$ and D. W. L. 
Sprung$^b$}
\address{$^a$Redeemer University College, Ancaster ON L9K 1J4, Canada} 
\address{$^b$Department of Physics and Astronomy, McMaster University, 
Hamilton ON L8S 4M1}
\author{J. Martorell }
\address{Departament d'Estructura i Constituents de la Materia, Facultat
F\'{\i}sica, University of Barcelona\\  Barcelona 08028, Spain}
\date{\today}
\maketitle

\begin{abstract}
Time delay in electron propagation through a finite periodic system 
such as a semiconductor superlattice is studied by direct numerical 
solution of the time-dependent Schr\"odinger equation. We compare 
systems with and without addition of an anti-reflection coating 
(ARC). With an ARC, the time delay is consistent with propagation at 
the Bloch velocity of the periodic system, which significantly reduces the 
time delay, in addition to increasing the transmissivity. 
 \end{abstract}
\narrowtext
]


\section{ Introduction} 
Electron transport in a layered semiconductor superlattice (SL) 
\cite{Wack02} can in many cases be 
treated as a one-dimensional finite periodic system. With as few as 
$N =$ five cells, a well-developed band structure ensues \cite{FPP}. 
Within each allowed band, where the Bloch phase $\phi$ increases by 
$\pi$, the transmission shows narrow peaks determined by $N\phi(E) = 
m \pi$, $m = 1, 2, \cdots N-1$. Between these peaks the transmission 
touches an envelope of the minima which also has a simple 
description. In the forbidden bands, the Bloch phase $\phi$ acquires 
an imaginary part; as a result the transmission goes rapidly to zero.  
Pacher et al. \cite{Pach01} have used this property to design an 
electron band-pass filter. Further, by adding a quarter-wave cell at 
each end of the periodic array \cite{PG03,PG04}, they were able to 
increase the average transmission within the band from about $25\%$ 
to about $75\%$.  

In this work we will study the time dependence of an electron 
{\it wavepacket }passing 
through a superlattice, comparing the situation with and without an 
anti-reflection coating (ARC). In a series of papers 
\cite{MSM02,MSM02b,SMM03,SMM04}, some of the present authors have 
shown how to design an ARC which gives optimal transmission within a 
given miniband, by adding suitably configured potential cells on each 
end of a periodic array. The design depends on the shape of the 
potential cells making up  the periodic array, but not their number. 
An $r$-cell ARC consists of $r$ distinct potential cells on each end 
\cite{SMM03}. In the simplest case the periodic array consists of 
reflection symmetric cells; the ARC can also be made reflection 
symmetric by using symmetric cells and reversing their order at the 
opposite end. 

In the transfer matrix formalism, the electron wave function at fixed 
position $x$ is represented by two amplitudes, which can be treated 
as a spinor. The upper component corresponds to right-moving and the 
lower to left-moving waves. Passing though an arbitrary potential 
cell is described by the action of a transfer matrix on this 
spinor. For reflection symmetric cells, the transfer matrix can be 
represented in the Kard form \cite{SMM03,SMM04,Kard57}, which 
involves just two real parameters at given energy.  One of these is 
the Bloch phase, and the second, the impedance parameter $\mu$, is 
approximately constant over the middle of an allowed band, but 
diverges at the band edges. Adopting the Kard parameterization allows 
one the benefit of a powerful analogy to the precession of an 
electron spin in a magnetic field. The magnetic field direction has 
polar angle $\theta$, where $\tan \theta/2 = \tanh \mu/2$. The angle 
of precession is twice the Bloch phase: $2\phi$. An electron moving 
to the right corresponds to spin-up along OZ, and left-moving waves 
to spin-down. Passing through $N$ identical cells therefore amounts 
to precession by angle $2N\phi$; when this is an integer $m$ multiple 
of $2\pi$, the final electron state will be the same as the initial 
plane wave, (except for a phase,) which gives perfect transmission. 

A Bloch state is one where the wave function on either side of the 
potential cell differs only by the Bloch phase $\phi$. This is 
analogous to an electron whose spin is aligned along the direction of 
the magnetic field. The action of an ARC can therefore be understood 
as taking an initial spin-up state and rotating it to lie along the 
field direction. That is, it converts a right-moving wave into a 
Bloch state of the periodic array. Passing through any number of 
cells only adds an overall phase $N\phi$, and the downstream ARC 
reverses the alignment back to the spin-up state when the resonant 
condition holds. 

Without an ARC, electrons which are transmitted do so via narrow 
resonances as mentioned above, so one expects a significant time 
delay. Resonant states and their characteristic time dependence 
\cite{KP38,Sgt39} are key parts of nuclear physics. Perhaps the 
definitive discussion of the time evolution of wave packets in the 
vicinity of resonant states is that of Rosenfeld \cite{Ros65}. When 
the ratio $\Gamma/E_r$ of resonance width to energy is small, and the 
wave packet is narrow in energy compared to $\Gamma$, there is a 
broad range of intermediate times over which the system exhibits 
exponential time dependence, and therefore significant time delay 
compared to free propagation. The work of More and Gerjuoy 
\cite{Mo71,MG73} should also be mentioned. While the 3D system differs in 
important ways from 1D, the same conclusion holds. The SL is 
an interesting system to study precisely because placing $N-1$ resonances 
within an allowed band produces resonances with widths the order of a few 
meV, satisfying one of Rosenfeld's criteria. 

With an ARC, propagation via a Bloch state should be much 
faster. To quantify this effect we have performed numerical solutions 
of the time-dependent Schr\"odinger equation (TDSE). 
The subject of time delay in passing though a barrier is a 
controversial one (see the reviews by Hauge and St{\o}vneng 
\cite{Hau89}, Leavens and Aers \cite{Leav96},and  de Carvalho and 
Nussenzveig \cite{CN02} for example) but it is not our purpose to 
debate the merits or demerits of the various definitions that have 
been used (phase time, dwell time, Larmor time, etc.) This controversy 
continues in the context of wanting to define a single time to 
characterize the process. Our approach is to solve the time dependent 
Schr\"odinger equation directly. This provides a direct means of 
comparing the two situations (with/without an ARC), while 
providing much more information than a single number.

\begin{figure}          
\leavevmode
\epsfxsize=8cm
\epsffile{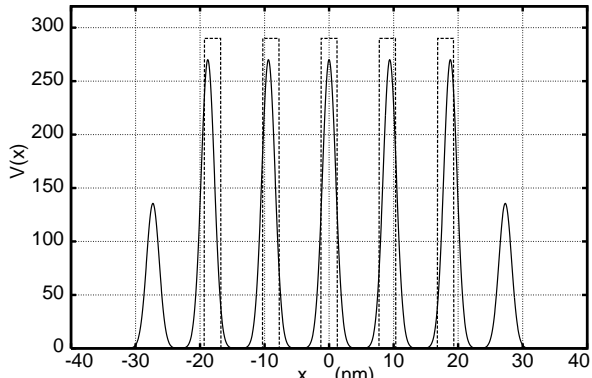}
\caption{Five cell array with a single-cell ARC, comparing the 
original square barrier cells to gaussian barrier cells.} 
 \label{fig1}
\end{figure}

    There have been several papers which discuss propagation in a 
superlattice {\it without} an ARC. Stamp and McIntosh \cite{SM96} 
made a careful study of the two-barrier case, which exhibits narrow 
quasi-bound state (QB) resonances. Their emphasis was on the 
role of the width (in energy) of the incident wave packet on exciting 
the quasi-bound states. They expanded the initial wave packet in 
stationary states of the scattering problem and propagated the 
solution in time by quadratures. This method is feasible for square 
barrier/well type potentials where the solutions are analytic. 
Pereyra \cite{SP03} used a similar approach for the superlattice 
problem. Bouchard and Luban \cite{BM95} considered electrons trapped 
in an infinite SL, in presence of an external electric field. Their 
emphasis was on the Wannier-Stark ladder of states localized by the 
applied field, and on finding Bloch oscillations. Their numerical 
method \cite{Koon86,GSS67} is similar to ours in using the implicit 
method to propagate the state forward in time.  However, they 
confined the system in a box and had to limit the total time so as to 
avoid reflection from the walls. In our work, transparent boundary 
conditions remove such restrictions. The work closest to ours is by 
Pacher and Gornik \cite{PG03,PG04}, who considered the effect of an 
ARC on transmission through a superlattice, following Pereyra for 
the time evolution. 

The numerical method we use was pioneered by Goldberg et al. 
\cite{GSS67}, and greatly improved by Moyer \cite{CAM04} who 
implemented the Numerov algorithm for the spatial variation and added 
transparent boundary conditions. The time-dependence is handled by 
the lowest-order Crank-Nicolson (implicit) method. The wave equation 
was followed for typically 15 ps, on a region (the ``system") $x_L < 
x < x_R$ of width $3$ to $8 \mu$m. Transparent boundary conditions 
were especially important in obtaining our results; they were applied 
at both ends of the system, so that as the wave function reaches 
those limits, it will proceed outwards without reflection. By 
integrating the outgoing flux at those boundaries, the reflection and 
transmission probabilities are accumulated. 

The initial state is a gaussian wave packet sufficiently wide in real 
space to correspond to a small uncertainty $\sigma_E$ in energy. For 
$\sigma_E/E \sim 3\%$ the root mean square (rms) width $\sigma_X$ is 
$140$ nm; for $\sigma_E/E \sim 0.4\%$ it is $1 \mu$m. The main 
requirement on the system width $x_R - x_L$ is to accommodate the 
initial wave packet, without overlapping the potential array.  
Narrower (in energy) packets would require a wider system and more 
computer time for the simulations. 

\section{The calculations} 
\subsection{Wave packet transmission}
The potential array is located on $a < x < b$ near the origin, very 
close to $x_R$. For this work we adapted a case previously studied. 
It corresponds to a GaAs/AlGaAs superlattice of Pacher et al., but 
with five barriers rather than the six of their device. Their 
modelling gave the barriers a height of approximately $290$ meV, and 
width $2.54$ nm, separated by wells of width $6.50$ nm. For 
convenience we used a constant effective mass $m^* = 0.071$ in all 
layers. Further, to simplify the numerical work, we replaced Pacher's 
square barriers by equivalent gaussian shaped barriers. 
 \begin{eqnarray}
V(x) &=& V_0\, e^{- x^2/(2 w^2)} , \quad -d/2 < x < d/2 \, .
\label{eq:td01}
\end{eqnarray}
The gaussian barrier has height $V_0 = 270.084$ meV and a width parameter $w 
= 1.02$ nm. The full cell width was set at $d = 9.4$ nm, only $4\%$ 
wider than the square barrier cell. These parameters were fitted to 
make the single-cell transfer matrices equivalent within the lowest 
allowed band, which runs from $50 < E < 74$ meV. The original and the 
gaussian potentials are shown in Fig. \ref{fig1}. 

\begin{figure}[htb]  
\begin{center}
\leavevmode
\epsfxsize=8cm
\begin{tabular}{cc} 
a) & \epsffile{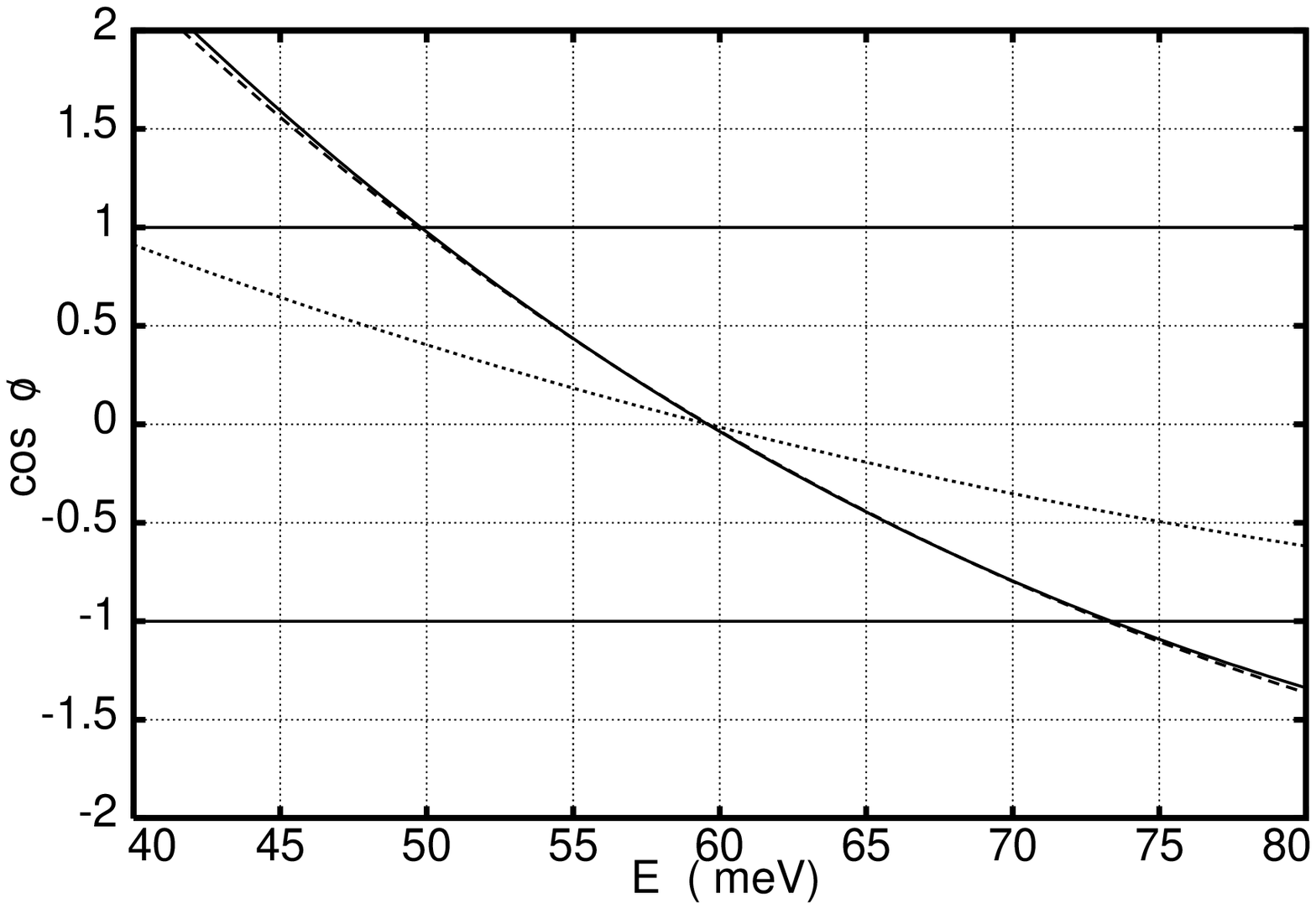} \\ 
b) & \epsfxsize=8cm \epsffile{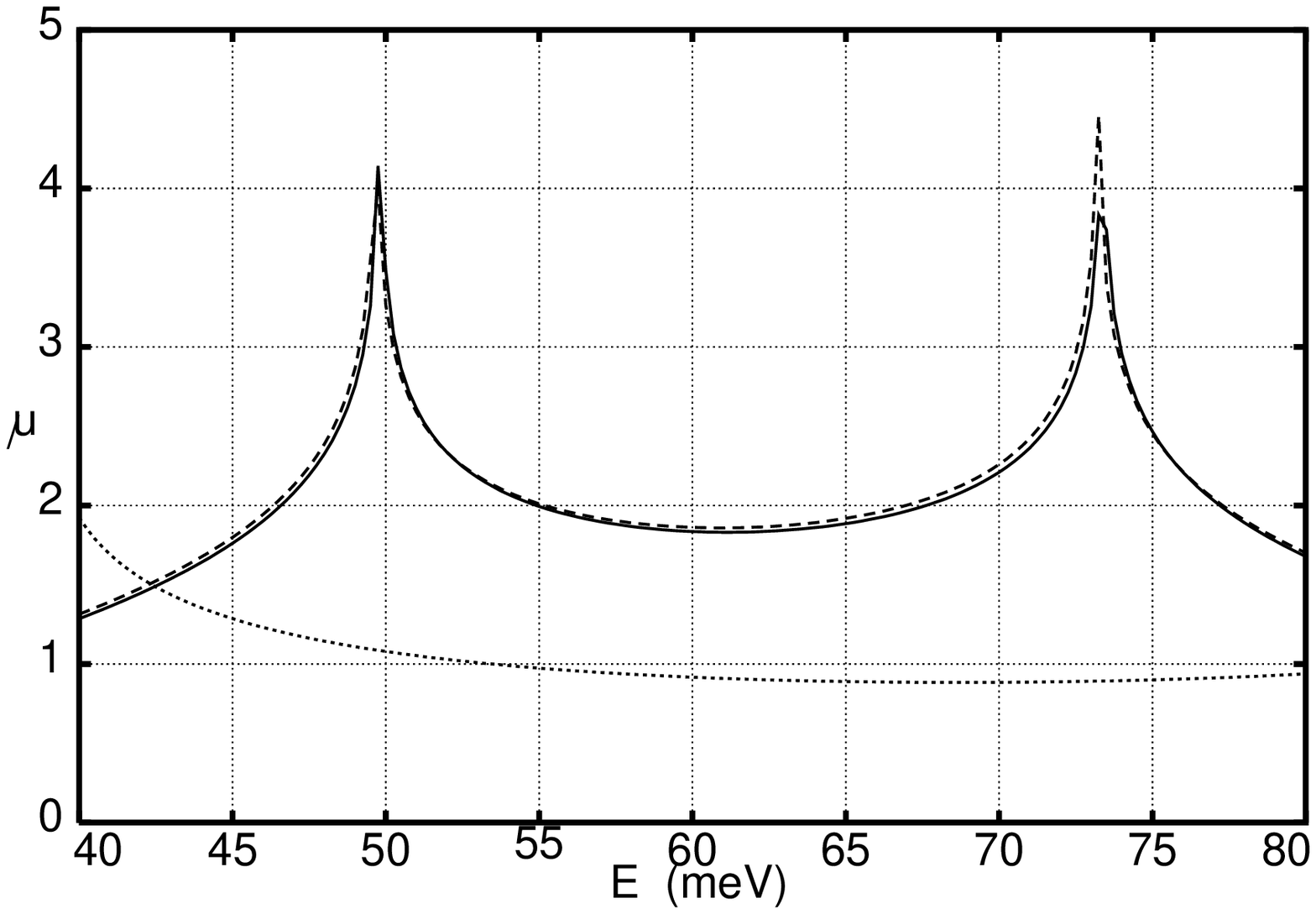} \\ 
\end{tabular}
\end{center}
\caption{Comparison of (a) Bloch phases of a square barrier cell 
(dash line) and our gaussian cell (solid line); also shown is 
$\cos \phi_A$ of a single-cell ARC (dotted line). 
(b) same for impedance parameters $\mu, \, \mu_A$.} 
\label{fig2}
\end{figure}

By equivalent, we mean that their scattering properties are 
accurately the same, across the allowed band of interest. Since the 
transfer matrix for a symmetric cell depends on only two Kard 
parameters, we need only fit these two, and the results are shown in 
Fig. \ref{fig2}. The $\cos \phi(E)$ are virtually identical for the 
original and the gaussian potential cell, as seen in Fig. 
\ref{fig2}(a). The third (dotted) line is the $\cos \phi_A$ of the 
optimal single-cell ARC. That gaussian has height $V_A = 135.64$ meV 
and width parameter $w_A = 0.98$ nm, with a total ARC cell width of 
$7.62$ nm. Also $\mu(E)$ was very close for both cells: see Fig. 
\ref{fig2}(b). (The lower line is $\mu_A$ for the ARC layer. For a 
single-cell ARC the prescription is $\mu_A = 0.5 \mu$ at the centre 
of the allowed band.) Having gaussian shape cells allowed us to use 
the Numerov method without having to worry about points of 
discontinuity of the potential. 

\begin{figure}                         
\leavevmode
\epsfxsize=8cm
\epsffile{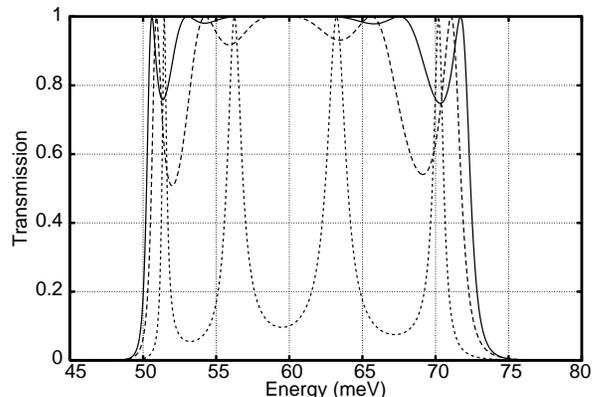}  
\caption{ Transmission for 5-cell array, bare and with single-cell 
and double-cell ARC (time-independent calculation).} 
\label{fig3}
\end{figure}

In  Fig. \ref{fig3} we show the transmission profiles of 
the 5-cell SL with no ARC (dotted line), a single-cell ARC (dashed line) 
and a two-cell ARC (solid line), for the gaussian cells array 
(computed using the usual time-independent methods). With the 
two-cell ARC, the high transmission region runs from 53 to 68 meV, 
and the satellite peaks are pushed closer to the band edges. This 
agrees well with the square barrier calculations in \cite{MSM02}. 

For solution of the time-dependent Schr\"odinger equation (TDSE) we 
chose gaussian wave packets. The initial state 
 \begin{eqnarray}
\psi(x,t=0) &=& \frac{1}{(2\pi \sigma_X^2)^{1/4}} \, e^{i k (x - x_0)} \, 
 e^{- (x-x_0)^2/(4  \sigma_X^2)} \,  
\label{eq:td02}
\end{eqnarray} 
is normalized to one particle. 
For a potential array sitting near the origin, the initial position $x_0$ 
has to be taken sufficiently negative so that the wave packet is well 
away from the periodic potential at time zero. For the more time consuming 
runs that followed the transmission across the entire band, a width 
$\sigma_X = 1000$ nm was used. When producing videos of the scattering 
process, a less demanding $\sigma_X = 140$ nm was chosen. 
Most of the work was done with $x_L = 
-7500$ nm to the left of the potential array, and a gaussian wave 
packet centered at the mid-point of that range. For the widest wave 
packets employed, the amplitude of the wave at $x_L$ and $0$ was 
therefore 0.0297 for $\sigma_X = 1 \mu$m, and was truncated to zero. 
The truncation introduces some high momentum components which have a 
small effect on time development, which is not visible in any of 
the graphs of this paper. Had we doubled the distances to 15 $\mu$m, 
the truncation would have been at 8 $\cdot  10^{-7}$ in amplitude and 
not at all discernible. The fractional standard deviation in energy 
can be written as 
 \begin{eqnarray}
\sigma_E/E &=& \frac{4.24 {\rm nm}}{\sigma_X} \,\sqrt{\frac{60\ {\rm meV}}{E}}~.
\label{eq:td03}
\end{eqnarray} 

\begin{figure}                         
\leavevmode
\epsfxsize=8cm 
\epsffile{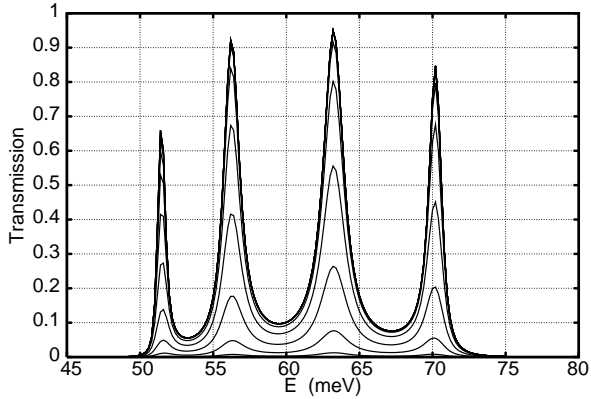}
\caption{Transmission for gaussian array without ARC showing time 
development. Contours are for  $t = 2\, (1.5)\, 14$ ps.} 
\label{fig4} 
\end{figure}

In Fig. \ref{fig4} we show transmission for the gaussian array 
computed using the TDSE. The right-moving flux at $x_R = x$ is 
 \begin{eqnarray}
j(x,t) &=& \frac{\hbar}{2 i m_e m^*} \bigr[ \psi^*(x,t) \frac{d 
\psi(x,t)}{dx} - \frac{d \psi^*(x,t)}{dx} \psi(x,t) \bigr] 
\label{eq:td04}
\end{eqnarray} 
and the integrated transmission is 
 \begin{eqnarray}
T(E_p,t) &=& \int_0^t j(x,t) dt  \quad {\rm with} \nonumber \\
T_{as}(E_p) &\equiv& \lim_{t \to \infty} T(E_p,t)~. 
\label{eq:td05}
\end{eqnarray} 
The label $E_p$ is mean energy of the wave packet. 
Corresponding expressions for the reflection probability $R(E_p,t)$ 
hold with the left-moving flux monitored at $x_L=x$. For 
packets narrow in energy $T_{as}$ should approach the transmission 
probability of the time-independent solutions, and this is seen 
to occur. 
The main difference with respect to Fig. \ref{fig3} is that the peaks 
are smeared by the finite width of the wave packet (approximately 
0.5 meV). Even so, they are so narrow that the tops are ragged due to 
the finite steps in $E_p$ of the wave packets used to generate the 
figure. (With finer steps and narrower (in energy) wave packets, 
convergence has been verified.) 

Also in Fig. \ref{fig4}, lines below the peaks show their build-up 
over time $T(E,t)$; this proceeds uniformly, with some bias to faster 
development at higher energies. This bias can be understood as a 
 simple velocity effect: the higher energy wave packets travel faster 
and reach the right wall $x_R$ sooner than the slower moving wave 
packets near the lower band edge. We note an apparent discrepancy with 
Fig. 4 of Romo \cite{Rom02}: at 0.6 ps he shows transmission 
in the troughs exceeding the asymptotic ($t \ge 60$ ps) transmission. 
However, as pointed out by an astuute reader, Romo's calculation is based 
on a very different initial state than ours \cite{MM52,GR97}. At time 
$t=0$ his initial state is a uniform standing wave confined by a 
mirror to the left of the array at $x < a$. The mirror is removed at 
$t=0$, and after an infinite time a new equilibrium state is achieved 
which is the stationary scattering state for a wave incident from the 
left. As the wave front advances to the right, the leading edge 
becomes smeared out, and oscillations develop behind, a process which 
Moshinsky \cite{MM52} called diffraction in time. Ultimately the 
limit as $t \to \infty$ of $|\psi(b,t)|^2$ should approach the time-
independent transmission probability $T(E)$. The curves at 
intermediate times, at a fixed position, show the passing of the wave 
front and the characteristic oscillations behind it. They do not 
represent the accumulation of $T(E_p,t)$ as in our calculation. 
Indeed, we can calculate using a long square wave packet normalized 
to one particle, and we find similar oscillatory behaviour in the 
build-up of both $|\psi(x,t)|^2$ and the flux, which is largely 
absent in the time-integrated flux. 

\begin{figure}                        
\leavevmode 
\epsfxsize=8cm 
\epsffile{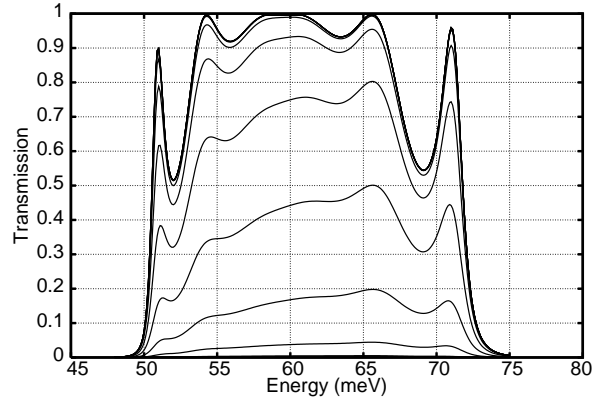} 
\caption{Transmission for gaussian array plus single-cell ARC, showing 
development over the same time intervals.} 
\label{fig5} 
\end{figure}

Fig. \ref{fig5} shows similar results for the same array plus a 
single-cell ARC. Again the asymptotic transmission is in good 
agreement with a time-independent calculation, but smeared by the 
finite width of the wave packet. Also, the build-up of the 
transmission profile proceeds smoothly with the above-noted velocity 
bias. 

 \begin{figure}                           
\leavevmode 
\epsfxsize=8cm 
\epsffile{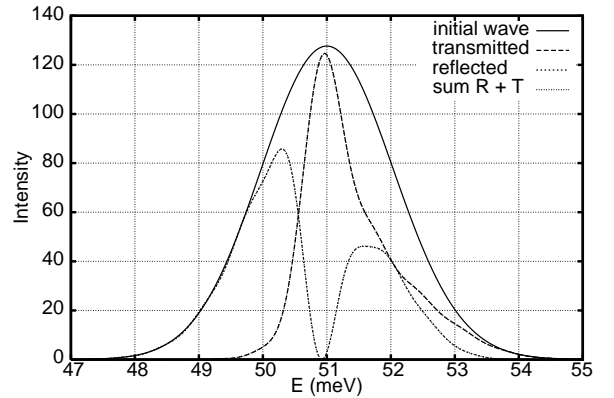} 
\caption{Energy content of the transmitted, reflected, 
initial and final wave packets near the 51 meV resonance.} 
\label{fig6} 
\end{figure} 

Near each band edge there is a satellite peak which resembles the 
narrow resonances of the periodic array. To study the character of 
these lines we scanned over them using spatially narrow wave packets. 
We enlarged the system to the right, so that there was room for both 
the reflected and transmitted wave packets to become well 
separated from the array. We then Fourier analysed the reflected 
and transmitted  wave packets individually. The Fourier transforms 
(moduli squared) 
$R(E)$ and $T(E)$ are plotted against energy in Fig. \ref{fig6} and 
the results are quite revealing. The T packet is very narrow compared 
to the incident wave, and centered just below 51 meV. The R packet in 
contrast has a node at this energy, and corresponds to the energies 
on either side. This shows that the transmission does proceed 
primarily through a narrow resonance. The transform (mod squared) 
of the complete  packet after scattering, denoted ``sum R+T" sits 
right on top of the curve for the initial wave packet, which is a 
testament to  the unitarity of the numerical work. 

\begin{figure}                           
\leavevmode 
\epsfxsize=8cm \epsffile{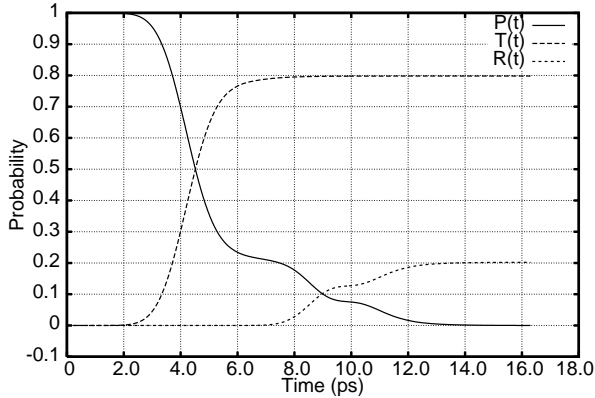} 
\caption{Accumulated transmission (T), reflection (R) and 
probability of remaining in the system (P) versus time, 
for a narrow wave packet at $E_p = 51 \pm 0.47$ meV.} 
\label{fig7} 
\end{figure}

The time-development of the reflection and transmission probabilities 
is shown in Fig. \ref{fig7} for a narrow (in energy) wave packet 
centered at $E_p = 51 \pm 0.47$ meV. $T(t)$ and $R(t)$ (see eq. 
\ref{eq:td05}) are accumulated 
as the waves exit from the right/left boundaries $x_R,x_L$. $T(t)$ 
begins to rise at 2 ps; this corresponds to the time taken for part 
of the initial packet to be transmitted and reach $x_R$. The longer 
time scale for $R(t)$ is due to the wider space on the left which 
accommodated the initial wave packet. One sees two steps in the 
$R(t)$ curve, one corresponding to direct reflection from the leading 
edge of the array, and the second one from entrapment before 
reflection.  The sum $R(t) + T(t) = 1 - P(t)$ is the complement of 
the probability $P(t)$ remaining within the system $(x_L,x_R)$: 
 \begin{eqnarray}
P(t) = \int_{x_L}^{x_R} |\psi(x,t)|^2 dx~. 
\label{eq:td06}
\end{eqnarray} 
(A 60 meV electron with effective mass $0.071$ travels at $545$ 
nm/ps; the time offsets are consistent with this.) 

\subsection{Quasi-Bound states}

\begin{figure}                         
\leavevmode 
\epsfxsize=8cm 
\epsffile{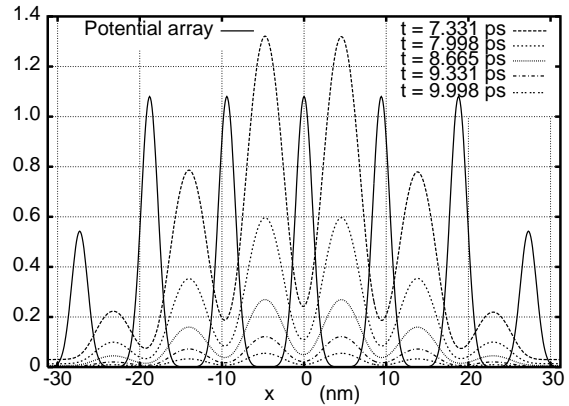} 
\caption{Decay of the QB state at 50.8 meV. The potential array 
with single-cell ARC (solid line) is shown for orientation.} 
\label{fig8} 
\end{figure}

At times near 6 ps $T(t)$ is approaching its asymptotic value 
$T_{as}$. 
Looking at the wave which is still trapped inside the array, one sees 
that it has a particular shape (shown in Fig. \ref{fig8}) and is decaying 
with mean life $\tau_n = 0.838 (2) $ ps. This corresponds to a 
half-width $\Gamma_n = 0.786$ meV. It implies a quasi bound state located at 
$E_n -i \Gamma_n/2$ in the lower half complex plane. Without the ARC, the 
position would be slightly displaced, corresponding to the shift of 
the outer peaks in Fig. \ref{fig3}.  

An electron inside the array sees four potential wells 
between the five stronger barriers. The QB states are built on the 
ground state in each well. Through coupling across the barriers 
four QB states are created, associated with the four transmission peaks seen 
in Fig. \ref{fig4} for example. Fig. \ref{fig8} shows that even with 
the ARC, the first  of these states persists (as does the last). Rosenfeld 
\cite{Ros65} discussed in detail the conditions under which the 
exponential decay of such states would be observed; the most 
important is that the ratio $\Gamma_n/E_n << 1$, which is well 
satisfied for our superlattice. Stamp and McIntosh \cite{SM96} 
emphasized another criterion, the ``interaction time" of the wave with 
the potential. This is basically the dwell time inside the potential. 
If it is too long compared to the lifetime of the QB state, the state 
will be decaying continuously as it is being fed, and the probability 
will not be built up to a significant extent. For the state of Fig. 
\ref{fig8}, $\tau_n$ is of the same order as the time delay shown in 
Fig. \ref{fig10}, below. 

The fourth QB state also survives the addition of the ARC, and 
implies a pole at $72 -i 0.767$ meV. The drawing would look very similar 
to Fig, \ref{fig8}, except that the wave function has nodes in 
each barrier. That is because the lowest state is nodeless, while the 
fourth state has alternating signs in the four wells.

\section{Time Delay}
\subsection{Results from time-dependent calculations}
The primary motivation for this work was to compare the time taken to 
traverse the array plus ARC, as compared to the simple array. However, 
in the case with ARC, the two satellite peaks appear to be narrow 
resonances of the same type as for the simple periodic array. Hence 
it is also of interest to compare transmission through  the satellite 
peaks with transmission via the Bloch states of the broad central 
maximum. 

Fig. \ref{fig7} shows that the build-up of the transmission peak at 
51 meV proceeds linearly in time, over the range $0.2 < T(t) < 0.6$. 
Similar linear behaviour applies throughout the allowed band. In the 
absence of a potential a similar curve is found, with some time 
displacement. By comparing the two, a reliable time delay (or 
advance) can be deduced which depends little on just which point is 
selected. For larger and smaller times the build-up is non-linear, 
and the converse is true.

\begin{figure}[htb]                      
\begin{center}
\leavevmode
\epsfxsize=8cm
\begin{tabular}{cc} 
a) & \epsffile{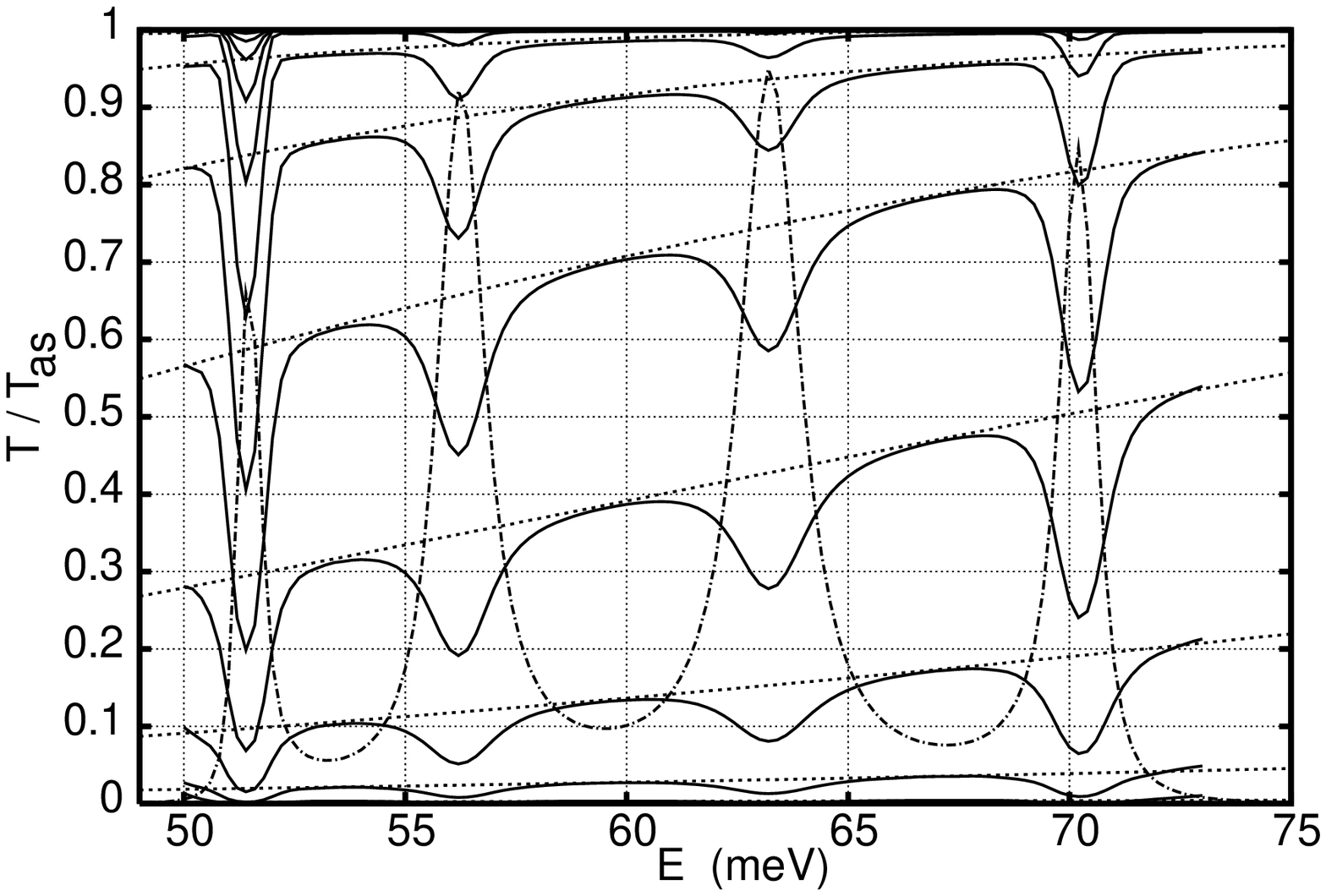} \\ 
b) & \epsfxsize=8cm \epsffile{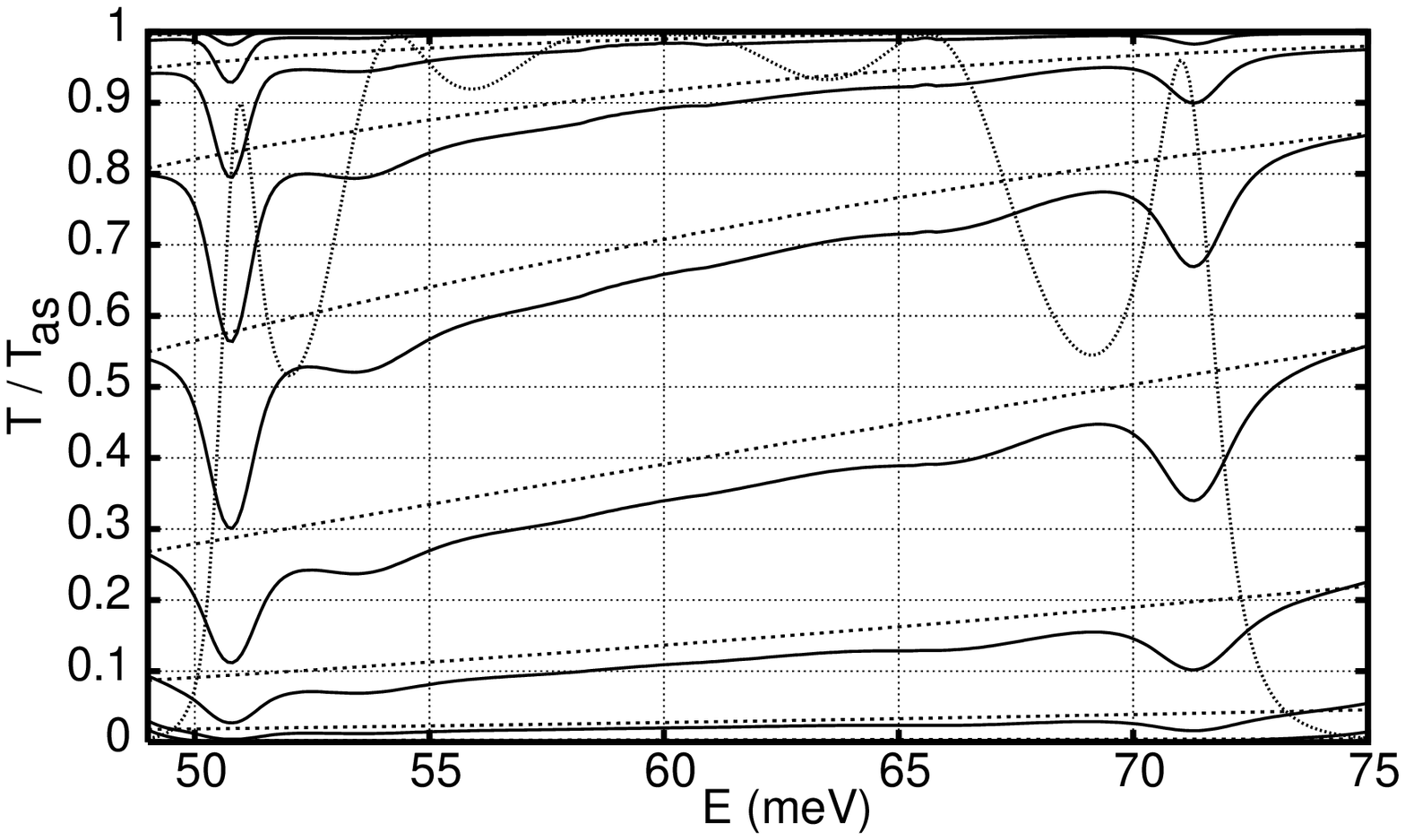} \\ 
\end{tabular}
\end{center}
\caption{Ratio $T/T_{as}(E)$ with (solid line) and without potential 
(dotted line) for (a) five barrier periodic system and (b) with 
single-layer ARC added. The contours are plotted at intervals
$t = 2 \, (1.5)\, 14$ ps. Also shown (dotted) is the transmission $T(E)$, 
for reference.} 
\label{fig9}
\end{figure}

In Fig. \ref{fig9} (a) we have plotted the ratio of $T(t)/T_{as}$ 
(solid line) as a function of energy across the whole band. The 
several lines correspond to different times. The inclined straight 
dotted lines are the same thing in the absence of any barriers (free 
propagation). In the background as a chain line, the transmission 
probability $T(E)$ is plotted for reference. Except at the four 
transmission peaks, the solid and dotted lines agree well. Also, the 
rate of build-up is the same as in the absence of the barriers. That 
says that in the transmission troughs, almost nothing is being 
transmitted but it does so with no time delay. Under the four peaks, 
the build-up though still linear in time, is significantly retarded 
compared to free propagation. The inclination of the dotted lines we 
interpret as a velocity effect: the higher energy electrons travel 
faster and arrive at $x_R$ sooner. 

To deduce a time delay, we take the difference between the solid and 
dotted lines, averaged over the range $0.3 < T(t)/T_{as} < 0.7$.  
This avoids using the non-linear portion of the $T(t)$ curve, as 
already discussed in connection with Fig. \ref{fig7}. It would make 
little difference had we simply taken the time delay at ratio 0.5 . 
The resulting time delay is plotted in Fig. \ref{fig10} (a). This 
procedure is consistent with the work of Dumont and Marchioro, who 
argue that for suitably chosen wave packets and by measuring the flux of 
particles at $x_R$, one can define a ``tunneling time probability 
distribution", which is basically the derivative of the integrated 
transmitted flux $T(t)$ plotted in Fig. \ref{fig7}, but normalized 
to its asymptotic value $T_{as}$. 

Fig. \ref{fig9} (b) shows the same ratio $T(t)/T_{as}$ after adding a 
single-cell ARC. Except at the two satellite peaks, the solid and 
dotted lines are roughly parallel, but not touching, showing that 
there is a delay for propagation via a Bloch state. The delay is 
significantly greater under the satellite peaks. The corresponding 
time delays are plotted in Fig. \ref{fig10} (b). Without an ARC, they 
go to zero (or even negative) in the transmission valleys, but at the 
peaks they range from 0.75 to 2.6 ps. With a single-layer ARC the 
delay is 0.25  ps over the middle of the band, jumping to 1.6 and 0.8 
ps under the satellite peaks. So, for those electrons that are 
transmitted, adding the ARC cuts the time delay in half, while 
greatly increasing the average transmissivity. [Also shown in Fig. 
\ref{fig10} (b) is a simple estimate of the time delay for traversing 
five cells at the Bloch velocity, ignoring any delay in the ARC 
cells. It can be seen that this estimate is the right size, and the 
downward trend from left to right also agrees. Because the Bloch 
velocity vanishes at a band edge, the curve diverges there. 
Calculations with ten and fifteen cells show that the satellite peaks 
move closer to the divergence at the band edge. The delay at 51 meV, 
1.6 ps, is about double the lifetime of the quasibound state, but at 
70 meV the two are about equal. Similar conclusions hold when a two-
cell ARC is added, but for brevity we do not include them here. 

\subsection{Comparison with time-independent calculations}

Time delay can also be computed in the time-independent formalism. 
From the vast literature on this subject, we base our 
discussion on relatively recent work of Nussenzveig \cite{Nuss00,CN02}. He 
argues in favour of the ``average dwell time" as the most appropriate 
measure of the time taken to pass a barrier, and lists its 
advantages. It is closely related to the ``phase time" or 
transmission group delay $\hbar d\tilde{\eta}/dE$ \cite{S60} originally 
introduced by Wigner and Eisenbud. Here $\tilde{\eta}$ is the phase of the 
complex transmission amplitude, closely related to the transfer 
matrix element by $M_{11} = 1/t$, as we now explain. 

In the transfer matrix method, our wave functions are defined with a 
different phase than generally used: namely the phase is set to zero on 
each side ($x=a,\, b$) of the potential array, rather than at the 
origin. Adopting Nussenzveig's notation for the asymptotic wave 
function to left and right, we compare our $\psi(x)$ with the usual 
convention $\tilde{\psi}(x)$ as follows: 
 \begin{eqnarray}
\psi(x) &\sim& [ e^{ik(x-a)} + r e^{-ik(x-a)} \, ; \, t e^{ik(x-b)} ] 
\nonumber \\  
\tilde{\psi}(x) &\sim& [\,\, e^{ikx} \,\,\,+\,\,\, \tilde{r} e^{-ikx} 
\,\,\, ; \,\, \qquad \tilde{t} e^{ikx} ] \nonumber \\  
e^{ika}\, \psi(x) &\sim& [e^{ikx} \,\, + \,\, r e^{-ik(x-2a)} \, ; 
\, t e^{ik(x-w)} 
]~, 
\label{eq:td07}
\end{eqnarray} 
where $w = b-a$ is the total width of the potential. It follows that 
the phase $\eta$ of our transmission amplitude is related to the 
usual one by $\tilde{\eta} = \eta - kw$.

Then the phase time delay is  
 \begin{eqnarray}
\tau_{ph} = \hbar \frac{d \tilde{\eta}}{d E} 
 &=& \hbar \Bigl[ \frac{d \eta}{d E} - w \frac{d k}{d E} \Bigr] 
\, , 
\label{eq:td08}
\end{eqnarray} 
\noindent 
where $\eta$ is the phase of our transmission amplitude and $w$ is 
the width of the superlattice (not the entire system) 
under consideration. Since $dE/dk$ is the 
velocity $v_f$ of a free particle, the contribution $w/v_f$ 
represents the time taken to cross the superlattice assuming zero 
potential.  In view of our earlier estimate for the free 
velocity, it is of the order of $0.1$ ps. Thus, $d\tilde{\eta}/dE$ 
gives phase time delay, while our $d\eta/dE$ gives phase time. 

For the case of unilateral incidence on a symmetric potential, which 
we use, Nussenzweig arrives at his eq. (20) for the mean dwell time 
delay as a spectral average over the phase time, plus an oscillatory 
term. In our notation, his result 
can be written as in eq. \ref{eq:ta03} in the appendix. 

In Fig. \ref{fig10} we show the time delay extracted from our 
time-dependent calculations, both before (a) and after (b) a single 
cell ARC has been added. In both panels, the lower dashed line would be the 
phase time delay, assuming that in the periodic array the wave 
propagates at the Bloch velocity of the infinite periodic system: 
 \begin{equation}
v_{Bl} = {- d \sin \phi \over \hbar} {{\partial E}\over {\partial\cos \phi}} 
\equiv d/\tau_{Bl}~,
\label{eq:td09}
\end{equation}
where $d$ is the cell size. Within an allowed band, $\cos \phi$ is 
generally quite linear in $E$. 

In Fig. \ref{fig10}(b) the dash-dot line includes an estimate of the 
time delay for passing through the ARC layers as well as the five 
central cells (it is a 10\% effect). This Bloch time-delay agrees 
quite well with the result of our time-dependent calculations, 
especially in its general trend. This confirms our 
understanding of the action of the ARC layers: they convert the 
incident plane wave state into a Bloch state of the periodic system. 
The divergence at the band edge arises because the Bloch velocity 
vanishes there. 

In Fig. \ref{fig10}(a) the dash-dot line is the locus of phase time 
delay at transmission maxima. 
While maximal time delays do not occur exactly at the 
transmission maxima, they do lie close together for this type of 
potential cell, and the difference would only be visible on a magnified 
drawing. 
The locus passes about 10\% above the 
peaks, which we ascribe to two effects: (i) the finite steps in mean 
wave packet energy in our calculations, which may miss the top, and 
(ii) the finite energy width of the wave packet which smears 
out the result based on plane waves, as in eq. \ref{eq:td10}. This is 
seen more clearly in Fig. \ref{fig11}. Interestingly, the time delay 
without an ARC oscillates around the Bloch time delay in Fig. 
\ref{fig10} (a); this is not particularly obvious from the present 
case of five cells, but if one compares systems with ten to fifteen 
cells it is quite striking.

 \begin{figure}[htb]      
\begin{center}
\leavevmode
\epsfxsize=8cm
\begin{tabular}{cc} 
a) & \epsffile{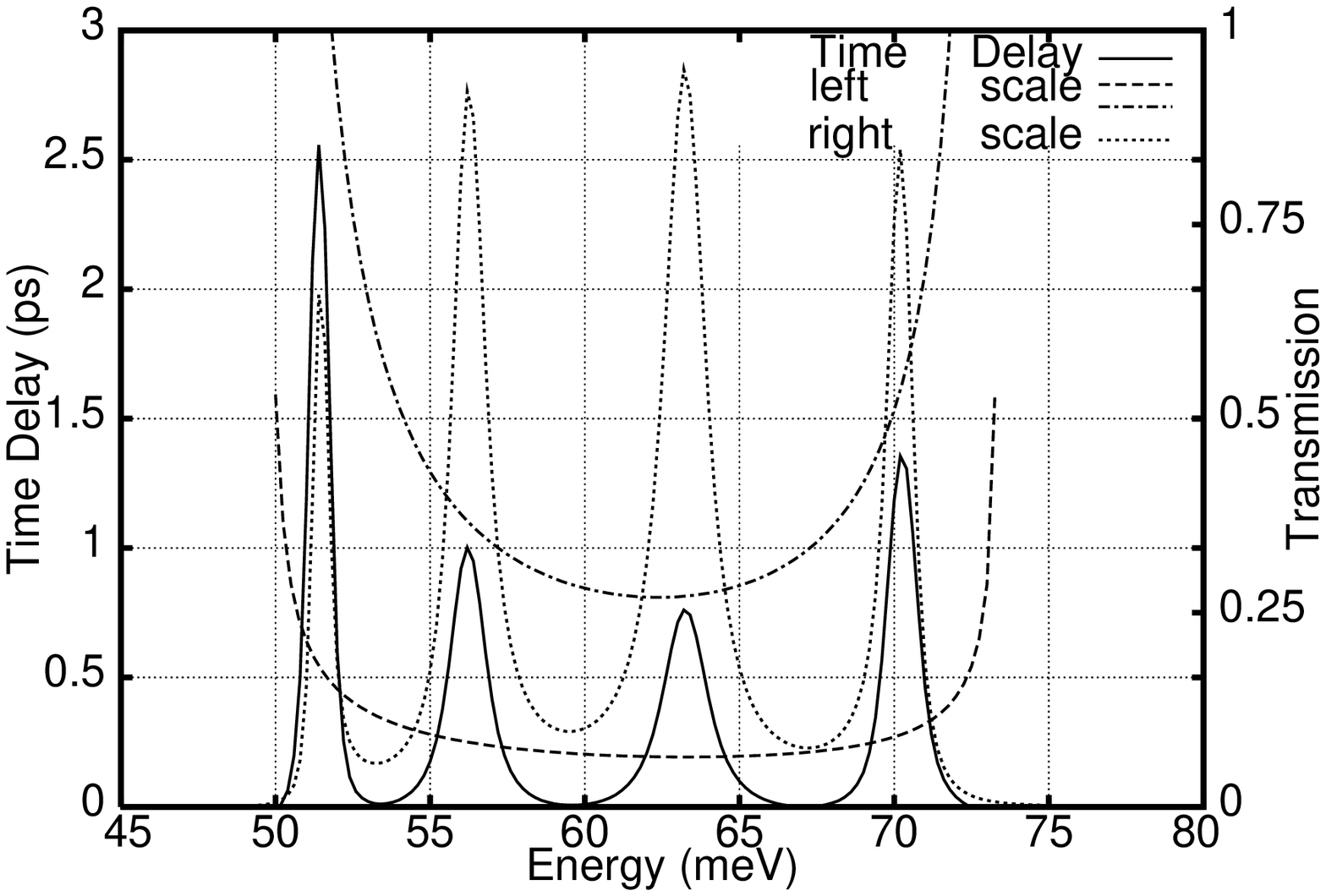} \\ 
b) & \epsfxsize=8cm \epsffile{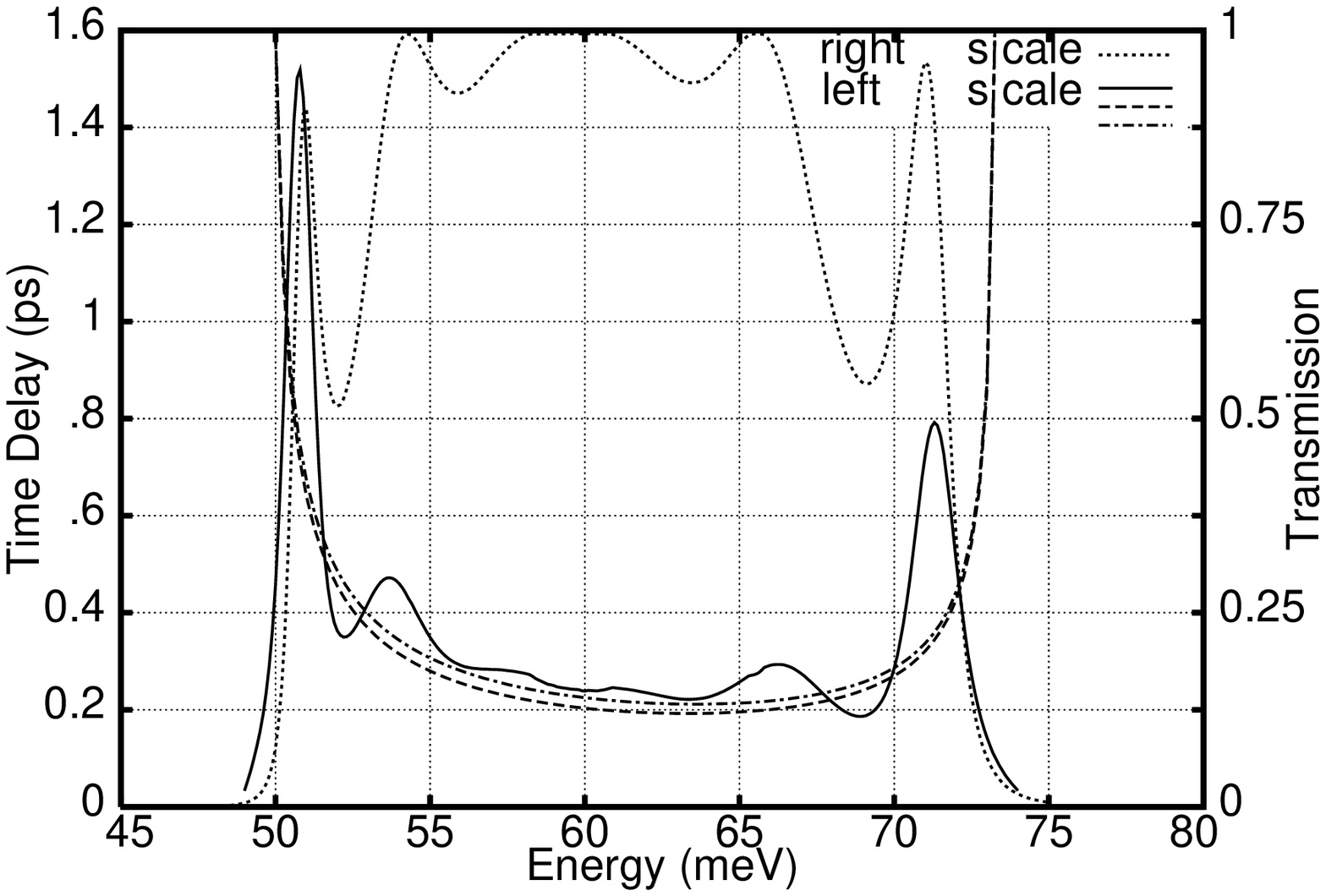} \\  
\end{tabular}
\end{center}
\caption{Time delay (solid line) 
for electrons transmitted through (a) the five 
barrier periodic system and also (b) with 
single-layer ARC added. Also shown is the transmission $T(E)$ 
(dotted line) for reference.} 
\label{fig10}
\end{figure}

The locus of time delay at transmission maxima is derived as follows: 
For a finite periodic system, we know that \cite{FPP} 
 \begin{equation}
{1\over t_N} = {1\over {\sin \phi}} \left( {1\over t} \sin N\phi - 
\sin (N-1) \phi\right) \ . 
 \label{eq:td10}
\end{equation}
We write $t = |t| e^{i\eta}$ and $t_N = |t_N| e^{i\eta_N}$ for $N$ 
cells.  Thus $\eta_N$ can be expressed in terms of $|t|,\,\eta$ and the 
Bloch phase $\phi$. Using eq. (\ref{eq:td08}) for the time delay we can 
similarly express $\tau$ in terms of the single cell parameters. To look 
for maxima, we set the derivative with respect to energy to be zero, 
and $\sin N\phi = 0$, because transmission maxima occur when $N\phi$ 
is an integer multiple of $\pi$ (see first paragraph of this paper.) 
The result is 
 \begin{equation}
{\hbar} \frac{d \eta_N}{d E}\Bigr|_{max} = \frac{ N d}{ |t| 
v_{Bl}} \, \frac{\sin \eta}{\sin \phi} 
= N \tau_{Bl} \cosh \mu~. 
 \label{eq:td11}
\end{equation}
where $\tau_{Bl}$ is the time taken to cross one cell at the Bloch 
velocity, and $\mu$ is the impedance parameter. The plot of $\cosh 
\mu$ in Fig. \ref{fig2} explains the shape and height of the curve 
immediately.    

From eq. \ref{eq:td11} we subtract the last term of eq. 
(\ref{eq:td08}), leading to an expression for the locus of 
time delay at transmission maxima: 
 \begin{equation}
 \tau_{loc} =  N \tau_{Bl}\, \cosh \mu - 
\frac{Nd}{v_{free}} \ . 
 \label{eq:td12}
\end{equation}

Results using eq. (\ref{eq:td08}) are in excellent agreement with those 
of the time-dependent calculation, when allowance is made for 
averaging over energy in the neighbourhood of the sharp resonances. 
There the finite width (in energy) of our incident 
wave packet mainly reduces the height of the peaks of the time-independent 
result, by 5 to 10\%, acording to eq. (\ref{eq:td13}). 
An example is shown in Fig. \ref{fig11}. 
In this drawing, we should have averaged the phase time over the 
spectral content of the initial state. We did not, but because our wave 
packets are narrow in energy, convolution only 
reduces the heights of the narrow peaks, which can be estimated from 
 \begin{eqnarray}
g(E) &\sim& \frac{f(E)}{1 + (2\sigma_E/\Gamma)^2\, f(E)}~. 
\label{eq:td13}
\end{eqnarray}
This assumes that the unsmeared function $f(E)$ is of Breit-Wigner 
form with width $\Gamma$, and is wide compared to the width in energy 
$\sigma_E \sim 0.4$ meV of the wave packet. Eq. (\ref{eq:td13}) agrees 
well with the reduction seen in Figs. \ref{fig10} and \ref{fig11},  
both for the time delay and transmission peaks. Mostly hidden below 
the solid line in Fig. \ref{fig11} is our time-delay from 
Fig.\ref{fig10} (a). One can see that except just below the peaks, the 
agreement is excellent.

Neither have we included the oscillatory term in the mean dwell time 
delay, which arises from interference between the incident wave and 
the reflected wave \cite{S60,Raz03}.  In our calculations there are 
indeed spectacular interference effects within and to the left of the 
potential array, as the reflected wave is generated. But after the 
reflected wave is well separated from the potential and reaches the 
counter at $x_L$, these oscillations have disappeared, and do not 
show up in the integrated flux. Neglect of this term, in the present 
calculation, is justified in the appendix.

Pacher and Gornik \cite{PG03,PG04} have also computed tunneling times 
using a similar formula following Pereyra \cite{P00}, with similar 
results. They did not perform  time-dependent 
calculations to establish the validity of the result. Furthermore, in 
our reading, Pereyra's derivation \cite{SP03} assumes that the reflection 
amplitude has a fixed modulus, and only the phase is varying, which 
is obviously questionable in the case of sharp resonances. 
After this work was submitted, Pacher et al. \cite{PBG05} have also 
discussed  phase time delay at the transmission maxima, deriving the result 
in eq. \ref{eq:td11} and many others.

\begin{figure}                      
\leavevmode 
\epsfxsize=8cm 
\epsffile{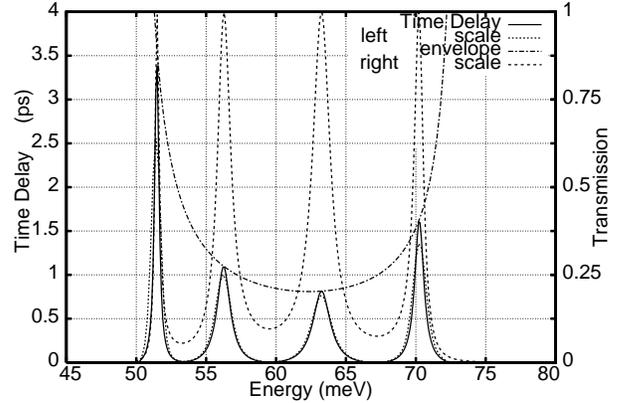} 
\caption{Time Delay (solid line) computed from phase of $t_N$, and 
(dashed line) from the time-dependent calculation, just visible below the 
peaks. The dash-dot line is the locus, eq. \protect{\ref{eq:td12}}. 
Also shown for reference is the transmission (dotted line). } 
\label{fig11} 
\end{figure} 

\section{Conclusion} 
Time dependence of scattering from a finite periodic potential array 
was studied by direct numerical solution of the Schr\"odinger 
equation. It was verified that the narrow transmission peaks are 
associated with quasi-bound states of the array. These resonances 
entail time delays of order 1 to 2 ps, while in the transmission 
minima the delay vanishes. Upon addition of 
an anti-reflection coating, the broad central transmission maximum 
corresponds to transmission via Bloch states, with a time delay of 
order 0.2 to 0.3 ps, as seen in Fig. \ref{fig10} (b). The satellite 
peaks near the band edge continue to proceed through QB states, but 
even their time delay is cut roughly in half, as compared to the bare 
periodic array. 

Our incident wave packets had widths in energy of order 0.4 meV; this 
was feasible due to the application of transparent boundary 
conditions at the edges of the region considered. A further 
improvement in the method of solution is possible, by going from 
first-order to third-order Crank-Nicolson integration for the time 
dependence \cite{PSV99,PSV00}. This would match the truncation error 
of the Numerov method, while greatly speeding up calculations. 
Systems under bias of an applied electric field \cite{MSM04} can also 
be handled by this method. Extension to two-dimensional systems is 
also under consideration. 

\acknowledgements

We are grateful to NSERC-Canada for Discovery Grants SAPIN-8672 
(WvD), RGPIN-3198 (DWLS) and a Summer Research Award through Redeemer 
University College (CNV); and to DGES-Spain for continued support 
through grants PB97-0915 and BFM2001-3710 (JM). We also thank Gigi 
Wong for assistance in redrawing Figs. 10 and 11, and R. S. Dumont for 
illuminating discussions.

\appendix

\section{Spectral average of the dwell time delay} 
For our wave packet eq. \ref{eq:td02}, the Fourier transform is 
 \begin{eqnarray}
\psi_0(q) &=& [ 8\pi \sigma_X^2]^{1/4} \, e^{-iqx_0} e^{- (q-k)^2 
\sigma^2}~,
\label{eq:ta01}
\end{eqnarray} 
normalized according to 
 \begin{eqnarray}
\int_{-\infty}^\infty \, |\psi_0(q)|^2 \frac{dq}{2\pi} = 1~. 
\label{eq:ta02}
\end{eqnarray} 
The spectral weight function is therefore $A(q) = |\psi_0(q)|^2 / 
(2\pi)$. 
Nussenzweig's eq. (20) for the time delay by a symmetric potential, 
with unilateral incidence, in our notation becomes 
 \begin{eqnarray}
&& <\Delta t_d>^{symm}_\to  = \int_{-\infty}^{\infty} |\psi_0(q)|^2 
\nonumber \\ 
&& \quad \qquad \times 
\left[ \hbar \frac{d\tilde{\eta}}{dE} - \frac{\hbar |r(q)|}{2E(q)} \cos(2 q 
x_L - \tilde{\eta}) \right] \,\, \frac{dq}{2\pi} ~. 
\label{eq:ta03}
\end{eqnarray} 

The terms in square brackets are the dwell time delay in the 
mono-energetic case. Razavy \cite{Raz03} for example derived them by 
following the method of Smith \cite{S60}, albeit with some typos in 
his eq. (18.19). (One has to note that for a symmetric potential his 
two phase shifts are related by $\eta = \delta + \pi/2$.)

As stated earlier, we used a wave packet with a width $\sigma_X$ of 
order 1000 nm. The width in energy is of order $0.4$ meV, which is 
small compared both to the width of the allowed band, $25$ meV, and 
even of the resonances of a finite periodic potential array, which 
are in the range of 2 to 5 meV depending on the position in the 
band. (The narrowest states are those crowded against the band edge.) 
It is reasonable to treat $A(q)$ as narrow 
compared to the width of the peaks in transmission. 

The first term, $d\eta/dE$, varies on the scale of the band-width divided by 
$N$, the number of cells, about 6 meV in our calculation.  The slope 
of $\eta$ is steepest at a resonance, where $\eta = m\pi$, and is 
minimal at the mid-points between resonances.  The phase shift $\eta$ 
varies smoothly compared to the wave packet, and that leads to the 
conclusion stated in eq. \ref{eq:td13}: the spectral average   mainly 
reduces the height of each transmission peak, leaving its position 
and width unchanged. 

The second, oscillatory term, (OT), includes a factor  $\cos (2 k x_L 
- \eta(k))$. Since our $x_L \sim -7500$ nm., this is a very high 
frequency oscillation, given that we set the boundary $x_L$ far to 
the left of the potential. The reflection amplitude $|r(k)|$ varies 
on the same scale as the phase shift. In doing the spectral average 
it is reasonable to treat the small prefactor $\hbar |r(k)|/E(k) \sim 10 
$ fs as slowly varying in 
comparison to the cosine.  The mean value of the OT can be estimated by 
the method of steepest descents, leading to 
 \begin{eqnarray}
&& <OT> \sim   -\frac{\hbar |r(k)|}{2E(k)} \cos ( 2 k x_L - \eta(k)) 
\,\, \sigma_X \nonumber \\ && \quad \times 
\,\, \sqrt{\frac{2}{\pi}}\,\,\int_{-\infty}^\infty \, 
\exp[( i(q-k) (2 x_L) - (q-k)^2 2 \sigma_X^2] dq \nonumber \\ 
&=& -\frac{\hbar |r(k)|}{2E(k)} \cos ( 2 k x_L - \eta(k)) \,\, 
\exp [{-(x_L/\sigma_X)^2/2}]~. 
\label{eq:ta04}
\end{eqnarray} 
By taking $x_L$ sufficiently far to the left, the spectral average 
can be made as small as we please. Mostly we used  $x_L/\sigma_X = 7.5$, 
so the exponential factor is $e^{-28}$, multiplying a 
term which is already small.

\end{document}